# Notes on Monotone Recognition in Multi-Valued Grids

Levon H. Aslanyan and Hasmik A. Sahakyan

Institute for Informatics and Automation Problems of NAS RA
e-mail: lasl@sci.am, hsahakyan@sci.am

**Abstract**

A novel method of monotone recognition based on the partitioning of the grid into discrete structures isomorphic to binary cubes (called "cube-split" technique) was proposed in our recent work, and a theoretical level description of two algorithms /algorithmic schemes/ solving this problem was also introduced. This paper provides implementation details of those algorithms, as well as focuses on the recognition of monotone binary functions with a small number of units.

## 1. Introduction

Let $\Xi_{m+1}^n$ denote the $n$-th Cartesian degree of the set $\Xi_{m+1} = \{0,1,\cdots,m\}$:

$$\Xi_{m+1}^n = \{(a_1,\cdots,a_n)|a_i \in \Xi_{m+1}, i = 1,\cdots,n\},$$

or, in other words, $\Xi_{m+1}^n$ is the set of vertices of the $n$-dimensional $(m+1)$-valued discrete grid. The total number of vertices of $\Xi_{m+1}^n$ is equal to $(m+1)^n$. We consider a component-wise partial order "$\leq$" on $\Xi_{m+1}^n$ defined in the following way: for arbitrary vertices $a = (a_1,\cdots,a_n)$ and $b = (b_1,\cdots,b_n)$ of $\Xi_{m+1}^n$, $a$ precedes $b$ ($a \leq b$) if and only if $a_i \leq b_i$ for $i = 1,\cdots,n$. Then, $(\Xi_{m+1}^n, \leq)$ is a partially ordered set; we will use its Hasse diagram for geometrical interpretations.

$f(x_1,x_2,\cdots,x_n): \Xi_{m+1}^n \to \{0,1\}$ is called a *binary function* defined on $\Xi_{m+1}^n$. We say that $f(x_1,x_2,\cdots,x_n)$ is a *monotone* function if for any two vertices $a, b$ of $\Xi_{m+1}^n$, $a \geq b$ implies: $f(a) \geq f(b)$. The vertices of $\Xi_{m+1}^n$, where $f$ takes the value "1", are called *units* of the function. The set of units of $f$ is usually denoted by $N_f$. The vertices of $\Xi_{m+1}^n$, where $f$ takes the value "0" are called *zeros* of the function. $a^1 \in \Xi_{m+1}^n$ is called a *lower unit* of $f$, if $f(a^1) = 1$ and $f(b) = 0$ for every $b \in \Xi_{m+1}^n$ less than $a^1$. $a^0 \in \Xi_{m+1}^n$ is called an *upper zero* of $f$, if $f(a^0) = 0$ and $f(b) = 1$ for every $b \in \Xi_{m+1}^n$ greater than $a^0$.

Fig. 1 demonstrates the Hasse diagram of $\Xi_5^3$, and a monotone binary function $f$ defined on $\Xi_5^3$. Highlighted vertices (4,4,4), (4,4,3), (4,3,4), (3,4,4), (4,4,2), (4,3,3), (3,4,3), (3,3,4), (2,4,4), (4,3,2), (3,3,3), (2,4,3), (1,4,4), (1,4,3) are units of the function, where (4,3,2), (3,3,3), and (1,4,3) are its lower units. The rest of vertices of $\Xi_5^3$ are zeros of the function, and (4,4,1), (4,2,4), (3,4,2),(2,3,4), and (0,4,4) are its upper zeros.

We consider the problem of query-based algorithmic identification/recognition of monotone binary functions defined on $\Xi_{m+1}^n$. This problem is initially investigated by V. Korobkov and V. Alekseev, but also, consecutively, by many other authors [1-4]. For $m = 1$, this is the case of ordinary monotone Boolean functions defined on the binary cube $E^n$, $E^n = \{(\alpha_1, \cdots, \alpha_n) | \alpha_i \in \{0,1\}, i = 1, \cdots, n\}$. Hansel's chain-splitting technique of $E^n$ [5] is a well-known effective tool for monotone Boolean function recognition. The outline of the algorithm is as follows: the set of vertices of the binary cube is partitioned into disjoint chains of different lengths (there are a total of $C_n^{\lfloor n/2 \rfloor}$ chains in the $n$-dimensional cube). A key property of the Hansel's chains is that once the function values are known for all the vertices in all the chains of length $k$, then the function values, inferable by monotonicity, are unknown for at most two vertices in each chain of the next length $k + 2$. The maximum number of queries to recognize the monotone Boolean function defined on the $n$-dimensional cube is $C_n^{\lfloor n/2 \rfloor} + C_n^{\lfloor n/2 \rfloor + 1}$.

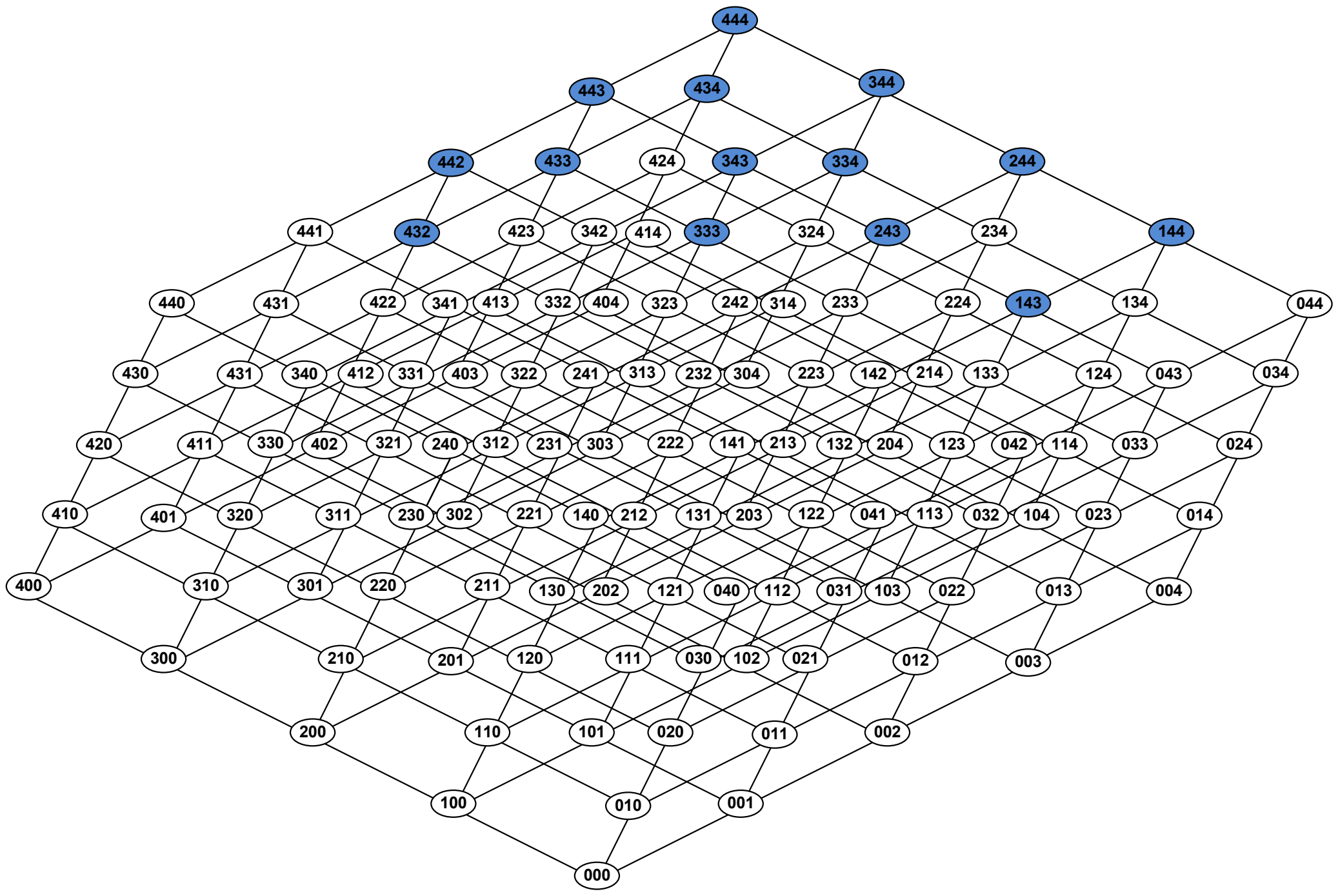

Fig. 1. Monotone function defined on $\Xi_5^3$.

An extension of this technique to the case of multi-valued grids and monotone binary functions is obtained in [2-3]. In [2], V. Alekseev developed the algorithm $U_0$ for recognition of a monotone binary function defined on $\Xi_{k_1 k_2 \cdots k_n} = \Xi_{k_1} \times \Xi_{k_2} \times \cdots \times \Xi_{k_n}$, ($\Xi_{k_i} = \{0,1, \cdots, k - 1\}, i = 1, \cdots, n$), which, in some sense, tries to generalize G.Hansel's algorithm. [2] proved that:

$$\frac{T(U_0)}{T(U_{opt})} \leq \frac{1}{2} \lceil log_2(k-1) \rceil,$$

Where $T(U_0)$ denotes the complexity of $U_0$, and $T(U_{opt})$ is the complexity of the optimal algorithm $U_{opt}$, $k = maxk_i$. It is also found that:

$$T(U_{opt}) \geq |M| + |N| \text{ and } T(U_0) \leq |M| + \lfloor log_2 k \rfloor \cdot |N|,$$

where $M$ and $N$ are the 2 sets of vertices in the middle layer area of $\Xi_{k_1 k_2 \cdots k_n}$:

$$M = \left\{(a_1, \cdots, a_n) \in \Xi_{k_1 k_2 \cdots k_n} : a_1 + \cdots + a_n = \left[\frac{1}{2}\sum_{i=1}^{n}(k_i - 1)\right]\right\},$$
$$N = \left\{(a_1, \cdots, a_n) \in \Xi_{k_1 k_2 \cdots k_n} : a_1 + \cdots + a_n = \left[\frac{1}{2}\sum_{i=1}^{n}(k_i - 1)\right] + 1\right\}.$$

In case of $\Xi_{m+1}^n$ the sets in the two middle layers are:

$$M_0 = \left\{(a_1, \cdots, a_n) \in \Xi_{m+1}^n : a_1 + \cdots + a_n = \left[\frac{m \cdot n}{2}\right]\right\},$$
$$N_0 = \left\{(a_1, \cdots, a_n) \in \Xi_{m+1}^n : a_1 + \cdots + a_n = \left[\frac{m \cdot n}{2}\right] + 1\right\},$$

and consequently, the estimate of complexity of the algorithm $U_0$ on $\Xi_{m+1}^n$ will be:

$$T(U_0) \leq |M_0| + \lfloor log_2 m \rfloor \cdot |N_0|.$$

A recent novel method of the monotone recognition based on a partitioning of the grid into discrete structures isomorphic to binary cubes (called "cube-split" technique) is proposed in [6], and two algorithms /algorithmic schemes/ solving this problem are also introduced. This paper provides implementation details of these algorithms, as well as focuses on the recognition of monotone binary functions with a small number of units.

The paper is organized as follows: Section 2 introduces the cube-splitting technique. Section 3 provides the implementation framework of these algorithmic schemes. Section 4 addresses some particular cases with a small number of units that comes from applications.

## 2. The Cube-Splitting Technique

In this section we introduce the cube-splitting technique of recognition of monotone binary functions [6]. Two homogeneous areas inside the $\Xi_{m+1}^n$ are defined in the following way:

- upper homogeneous area $\widehat{H}$, - this is the set of all "upper" elements of $\Xi_{m+1}^n$, i.e., elements with all-coordinate values $\geq m/2$;
- lower homogeneous area $\breve{H}$, – this is the set of all "lower" elements, i.e., elements with all-coordinate values $\leq m/2$. It is clear that:

$$|\widehat{H}| = |\breve{H}| = \begin{cases} (\frac{m+1}{2})^n \text{for odd } m, \\ (\frac{m}{2} + 1)^n \text{ for even } m \end{cases}.$$

The following results were introduced in [6].

(1) $\Xi_{m+1}^n$ can be split into $|\widehat{H}|$ disjoint discrete structures isomorphic to binary cubes:

$$\Xi_{m+1}^{n} = \mathcal{E}_1 \cup \cdots \cup \mathcal{E}_{|\widehat{H}|},$$

where every $\mathcal{E}_i$ contains exactly one vertex from $\widehat{H}$, while the remaining vertices of $\mathcal{E}_i$ can be determined uniquely by this vertex through the complementarity interchanges of the coordinate values. $\mathcal{E}_i \cap \mathcal{E}_j = \emptyset$, if $i \neq j$. The procedure is called "cube-splitting" of $\Xi_{m+1}^{n}$.

(2) The "cube-splitting" of $\Xi_{m+1}^{n}$ keeps the monotonicity property in the following way: let $F$ be a monotone binary function defined on $\Xi_{m+1}^{n}$, then either $N_F \cap \mathcal{E}_i$ is empty, or it satisfies the binary monotonicity property, i.e., for arbitrary vertex $a$ of $N_F \cap \mathcal{E}_i$, all vertices of $\mathcal{E}_i$ greater than $a$, also belong to $N_F \cap \mathcal{E}_i$ (for $i = 1, \cdots, n$).

By integrating (1) and (2), a novel monotone recognition method has been proposed in [6].

## 2.1 Definitions/descriptions

For each vertex $V_i = (v_{i_1}, \cdots, v_{i_n})$ of $\widehat{H}$ we compose the following set:

$$\mathcal{E}_{V_i} = \{(a_1, \cdots, a_n) \in \Xi_{m+1}^{n} | a_j \in \{v_{i_j}, m - v_{i_j}\} \text{ for all } j, 1 \leq j \leq n\},$$

and call $\mathcal{E}_{V_i}$ the *vertical equivalence class* of $V_i$.

$\mathcal{E}_{V_i}$ contains a unique vertex from $\widehat{H}$, - this is the vertex with all coordinates $\geq m/2$; and contains a unique vertex from $\breve{H}$, - this is the vertex with all coordinates $\leq m/2$. The remaining vertices of $\mathcal{E}_{V_i}$ can be obtained by component value inversions (with respect to $m$). $\mathcal{E}_{V_i} \cap \mathcal{E}_{V_j} = \emptyset$ for different vertices $V_i$ and $V_j$ of $\widehat{H}$. In this manner $\Xi_{m+1}^{n}$ can be split into $|\widehat{H}|$ disjoint sets /equivalence classes/ uniquely defined by the elements of $\widehat{H}$ (or $\breve{H}$).

The number of elements of $\mathcal{E}_{V_i}$ varies between $2^0$ and $2^n$ depending on the number of components of $V_i$ differing from $m/2$. Indeed, if $k$ denotes the number of components of $V_i$ differing from $m/2$, i.e. $k = \left|\{v_{i_j} | v_{i_j} \neq (m - v_{i_j})\}\right|$, then $|\mathcal{E}_{V_i}| = 2^k$. Notice that $k = n$ always for odd $m$. For example, Fig. 2 demonstrates $\mathcal{E}_{(3,4,3)}$, $\mathcal{E}_{(2,3,4)}$ and $\mathcal{E}_{(4,2,2)}$ in $\Xi_5^3$.

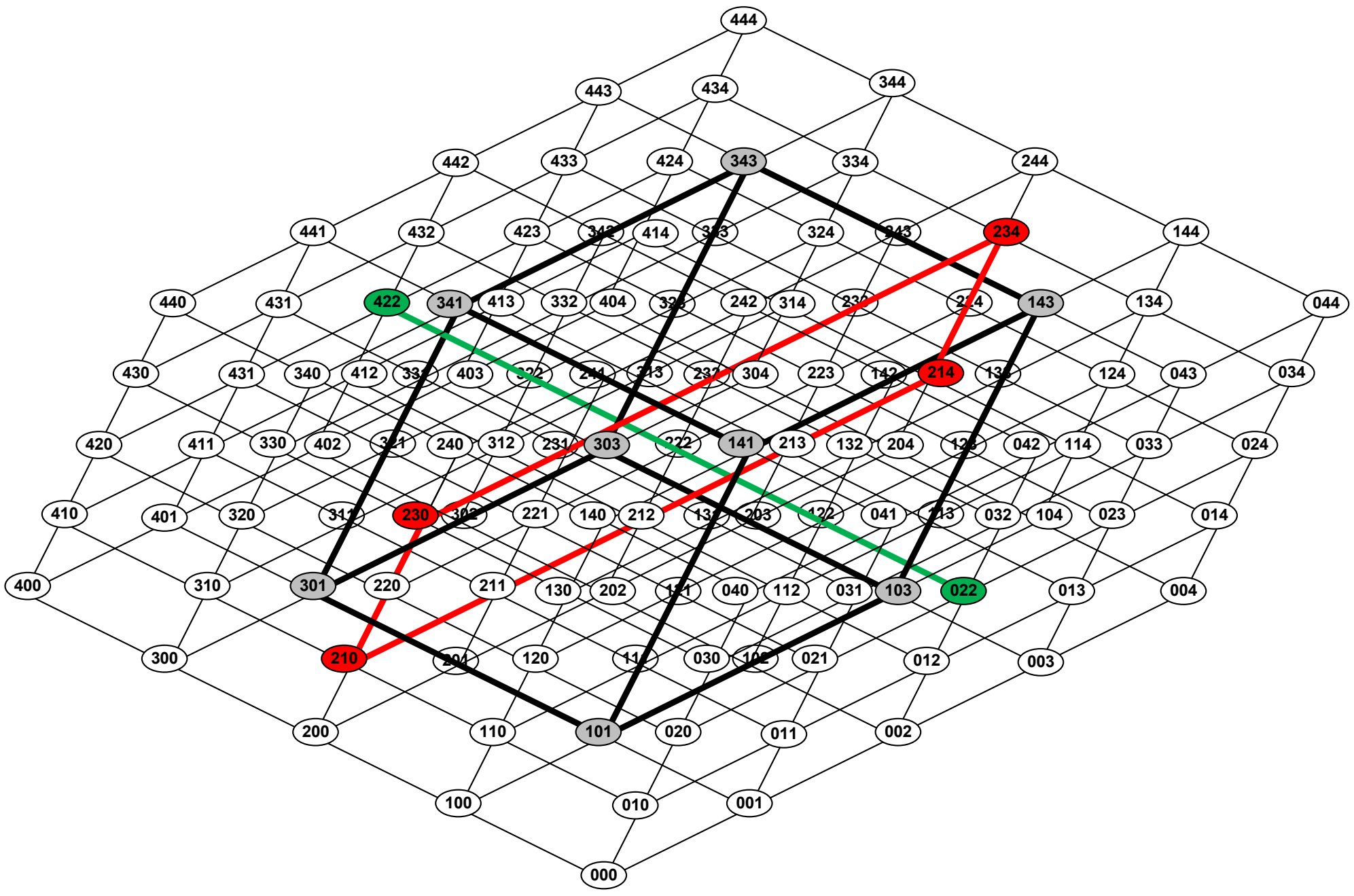

Fig. 2. Examples of cubes in a cube split of $\Xi_5^3$.

For every vertex $(a_1, \cdots, a_n)$ of the equivalence class $\mathcal{E}_{V_i}$, we distinguish its sub-list of all coordinates accepting a value differing from $m/2$, let this be the list: $(a_{s_1}, \cdots, a_{s_k})$. This list exactly fits the list of all coordinates of $V_i$ that are different from $m/2$. The reminder part of coordinates accepts the only value $m/2$ over the $V_i$, as well as over the whole set of vertices of $\mathcal{E}_{V_i}$. Now, identifying $(a_{s_1}, \cdots, a_{s_k})$ with the binary sequence $\beta = (\beta_1, \cdots, \beta_k)$ of length $k$ such that $\beta_j = 1$ if and only if $a_{s_j} > m/2$, - we map $(a_1, \cdots, a_n)$ into the vertex $\beta = (\beta_1, \cdots, \beta_k)$ of the $k$-dimensional binary cube $E^k$. In this manner, we obtain a 1-1 mapping $M: \mathcal{E}_{V_i} \longrightarrow E^k$. The vertex of $\mathcal{E}_{V_i}$ with all coordinates $< m/2$ is mapped into the vertex $(0, \cdots, 0)$ of $E^k$ (on the 0-th layer); the vertex of $\mathcal{E}_{V_i}$ with all coordinates $> m/2$ is mapped into the vertex $(1, \cdots, 1)$ of $E^k$ (on the $n$-th layer); and, in general, all vertices of $\mathcal{E}_{V_i}$, which have $l$ coordinates $> m/2$ (consequently, $m - l$ coordinates $< m/2$), are mapped into the vertices of $l$-th layer of $E^k$.

Hereafter, all structures (vertices, chains, cubes, functions, etc.) in $\Xi_{m+1}^n$ will be referred to as *origin*; and all structures (vertices, chains, cubes, functions, etc.) in binary cubes will be referred to as *induced*.

For example, the induced binary cubes for $\mathcal{E}_{(3,4,3)}$, $\mathcal{E}_{(2,3,4)}$ and $\mathcal{E}_{(4,2,2)}$, are given in Figure 3, (a), (b), and (c), correspondingly.

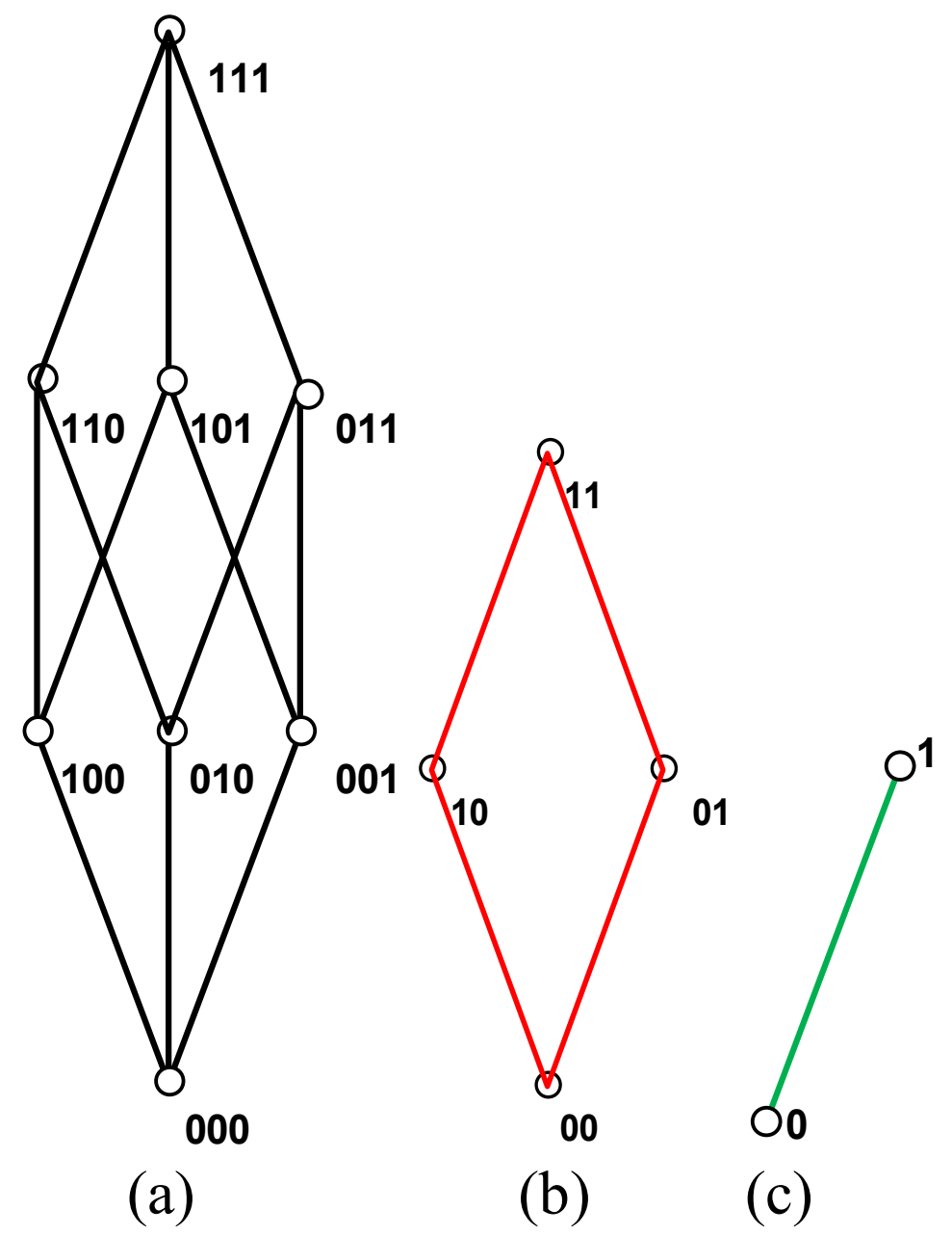

Fig. 3.

## 2.2 The Algorithmic Framework

Let $F$ be a monotone binary function (which should be recognized with the help of an oracle), defined on $\Xi_{m+1}^n$, and let $N_F$ denote its set of units of function $F$.

**Algorithm 1**

In a theoretical level description, the algorithm implements the following steps:

1. Apply the cube-splitting on $\Xi_{m+1}^n$ and let $\mathcal{E}_1, \cdots, \mathcal{E}_{|\bar{H}|}$ be the equivalence classes of the upper homogeneous elements.

2. Compose the corresponding induced binary cube $E_i$ for every $\mathcal{E}_i$.
3. Apply the Hansel's algorithm to recognize the induced Boolean function $f_i$, defined on $E_i$ as follows: for every $\beta \in E_i$, $f_i(\beta) = 1$ if and only if $F(b) = 1$, where $b$ is the origin of $\beta$ in $\Xi_{m+1}^n$ (for $i = 1, \cdots, |\widehat{H}|$).
4. Transfer the recognition results into the $\Xi_{m+1}^n$.

# 3. Implementation

## 3.1 Implementation details of Step 1 and Step2

We consider the lexicographic ordering of the vertices of $|\widehat{H}|$, where the smallest numerical values of coordinates are coming first. Thus, the smallest vertex of $\widehat{H}$ in this ordering is $(\frac{m+1}{2}, \frac{m+1}{2}, \cdots, \frac{m+1}{2})$ if $m$ is odd, and it is the vertex$(\frac{m}{2}, \frac{m}{2}, \cdots, \frac{m}{2})$, if $m$ is even; the greatest vertex is $(m, m, \cdots, m)$. Henceforth, we will assume that $\widehat{H} = \{V_1, V_2, \cdots, V_{\widehat{H}}\}$ is the lexicographically ordered set of upper homogeneous elements.

The cube splitting of $\Xi_{m+1}^n$ assumes that we compose for every vertex $V_i = (v_{i_1}, \cdots, v_{i_n})$ of $\widehat{H}$ its vertical equivalence class $\mathcal{E}_{V_i}$, and the corresponding induced binary cube to this.

But at this point we do not need to compose and keep (and further to map to the binary cube) the whole set $\mathcal{E}_{V_i}$; instead, with every vertex $V_i = (v_{i_1}, \cdots, v_{i_n})$ of $\widehat{H}$ we will keep the following parameters:

- the number $\tau_{V_i}$ of coordinates of $V_i$ differing from $m/2$, - this will determine the size of the induced binary cube $E_i$. When $V_i$, that is the issue, is evident, we will just use the notion $\tau$ for this,

- the positions of coordinates differing from $m/2$, we denote it by the vector $V_{i\neq}$, and

- the values of coordinates differing from $m/2$, we denote as the vector $V_{i\#}$.

$\tau$, $V_{i\neq}$, and $V_{i\#}$ will allow the easy reverse mapping, $RM : E^\tau \longrightarrow \mathcal{E}_{V_i}$, i.e., will allow to recover $\mathcal{E}_{V_i}$.

For example, with the vertex (2,3,4) of $\Xi_5^3$ we keep:

- numerical value 2, - this is the number of its coordinates differing from $m/2$,
- indexes 2,3 - these are the coordinate indexes, where the values are differing from $m/2$,
- and 3,4 are the values at the coordinates 2 and 3.

$\mathcal{E}_{(2,3,4)}$ is mapped into the 2-dimensional binary cube $E^2$ according to 2nd and 3rd coordinates of $(2,3,4)$, and the accompanying vectors are $V_{i\neq} = (2,3)$ and $V_{i\#} = (3,4)$.

The reverse mapping is as follows: given the pair $(2,3,4)$ and $E^2$ or alternatively, $V_{i\neq}$, $V_{i\#}$ and $E^2$. Consider an arbitrary vertex of $E^2$, for example, let $\tilde{\beta} = (1,0)$, then it follows that the origin of $\tilde{\beta}$ in $\mathcal{E}_{(2,3,4)}$ is $\tilde{\alpha} = (2,3,0)$, because:

- the first component, missing at $V_{i\neq}$, must be $m/2$, that is, $a_1 = 2$,
- the second component should not be inverted in accord to $\tilde{\beta} = (1,0)$, and, thus, $a_2 = 3$,
- the third component should be inverted, and thus, $a_3 = 4 - 4 = 0$.

In general, let $(\beta_1, \cdots, \beta_k)$ be an arbitrary vertex of the $k$-dimensional binary cube $E^k$, and let $(v_1, \cdots, v_n)$ be the upper homogeneous vector of the origin of $E^k$, and suppose that $s_1, \cdots, s_k$ are its coordinates differing from $m/2$. Then, the origin of $(\beta_1, \cdots, \beta_k)$ is $(a_1, \cdots, a_n)$, where:

$$a_{s_j} = \begin{cases} v_{s_j} \, if \, \beta_j = 1 \\ m - v_{s_j} if \, \beta_j = 0 \end{cases} \quad \text{for } j = 1, \cdots, k \tag{1}$$
$$a_i = m/2 \text{ for } i \neq s_1, \cdots, s_k.$$

## 3.2 Implementation Details of Step 3 and Step 4

In this part, Algorithm 1 recognizes monotone Boolean functions in the $|\widehat{H}|$ number of binary cubes of different sizes (some of the functions might be identically 0, but we do not know this fact beforehand), by applying the Hansel's algorithm.

Also at this step, we will not deal with the binary cubes themselves, and we will be using the chain algebras, and therefore, we have to map (by the reverse mapping) all induced structures (vertices, functions, chains, etc.) into their origins in $\Xi_{m+1}^n$.

For example, consider some monotone Boolean function $f_i$ on a $k$-dimensional binary cube. If we obtain the value $f_i(\tilde{\beta})$ on some vertex $\tilde{\beta}$ (in the process of the Hansel's algorithm), and know its origin upper homogeneous vertex $(v_1, \cdots, v_n)$ and also its coordinates differing from $m/2$ (let they be $s_1, \cdots, s_k$), then we can set that $F(\tilde{a}) = f_i(\tilde{\beta})$, where the coordinate values of $\tilde{a}$ are defined in accord to (1).

Similarly, we can map chains from the induced binary cubes into the $\Xi_{m+1}^n$.

For example, the maximum length chain <(000), (100), (110), (111)> in the cube in Figure 2 (a) (which is induced to $\mathcal{E}_{(3,4,3)}$), is mapped to the origin chain <(101), (301), (341), (343)> in $\Xi_{m+1}^n$.

Upon receipt of the oracle's response for a given vertex of the binary cube - the response is mapped into the origin vertex of $\Xi_{m+1}^n$. Certainly, the response value could also be extended by the monotonicity property to the other relevant vertices of $\Xi_{m+1}^n$. But in this research we prefer and emphasize the opportunity of the parallel implementation of the recognition algorithms in all the induced binary cubes, and so we keep them as separate nonintersecting processes.

# 4. Small Number of Units

Note that Algorithm 1 is worth applying when a large number of unit vertices of the monotone function appear in the upper homogeneous area $\widehat{H}$.

However, in some cases, mostly coming from applications, the function to be recognized has a small number of unit vertices in the upper area $\widehat{H}$, or its complement has a small number of zero vertices in the lower homogeneous area $\check{H}$. In the latter case we can recognize the complement of the monotone function. For this reason, the following points will be taken into account:

- $\mathcal{E}_{V_i}$ can be defined for each vertex of $\check{H}$ (obviously we will obtain the same set). In the example given in Figure 2 the highlighted sets of $\check{H}$ will demonstrate $\mathcal{E}_{(1,0,1)}$, $\mathcal{E}_{(2,1,0)}$ and $\mathcal{E}_{(0,2,2)}$, as well.
- The mapping $M: \mathcal{E}_{V_i} \longrightarrow E^k$ will be defined as follows: for every vertex $(a_1, \cdots, a_n)$ of $\mathcal{E}_{V_i}$ let $(a_{s_1}, \cdots, a_{s_k})$ denote its subsequence with all coordinates differing from $m/2$. Identify $(a_{s_1}, \cdots, a_{s_k})$ with the binary sequence $\beta = (\beta_1, \cdots, \beta_k)$ of length $k$ such that $\beta_j = 1$ if and only if $a_{s_j} < m/2$. The vertex of $\mathcal{E}_{V_i}$ with all coordinate values $< m/2$ is mapped into the vertex $(1,\cdots,1)$ of $E^k$, - this is on the $n$-th layer; the vertex of $\mathcal{E}_{V_i}$ with all coordinate values $< m/2$ is mapped into the vertex $(0,\cdots,0)$ of $E^k$, - this is on the 0-th layer; and, in general, all vertices of $\mathcal{E}_{V_i}$ which have $l$ coordinates $< m/2$ (consequently, $m-l$ coordinates $> m/2$) are mapped into the vertices of $l$-th layer of $E^k$.

- The reverse mapping: $RM: E_i \longrightarrow \mathcal{E}_{V_i}$ will be implemented as follows: let $(\beta_1,\cdots,\beta_k)$ be an arbitrary vertex of the $k$-dimensional binary cube $E^k$, and let $(v_1,\cdots,v_n)$ be the lower homogeneous vector of the origin of $E^k$, where $s_1,\cdots,s_k$ are coordinates differing from $m/2$.

Then the origin of $(\beta_1,\cdots,\beta_k)$ is $(a_1,\cdots,a_n)$, where:

$$a_{s_j} = \begin{cases} v_{s_j}\ if\ \beta_j = 0 \\ m - v_{s_j} if\ \beta_j = 1 \end{cases} \quad \text{for } j = 1,\cdots,k \qquad (2)$$
$$a_i = m/2 \text{ for } i \neq s_1,\cdots,s_k.$$

## 4.1 Constraints

Consider the application, which is the generalized model of the known association rule mining - in case where in addition to the presence or absence of elements in itemsets, the number of their repetitions is also included. The details are given in [7]. Here we highlight the following constraints/restrictions that may appear with this problem. In terminology of supermarket basket analysis, here we distinguish two postulations: some item exists in the current basket, and second, which is the actual number of that item in the basket.

Let $a_i$ be the repetition number of the $i$-th element, for $i = 1,\cdots,n$.

(1) the classic case is the (0,1) vector of item indicators in baskets, basket inventory.

(2) $a_1 + \cdots + a_n \leq r$, - the summary number of elements' repetitions (the basket volume) is restricted by $r$,

(3) $a_i \leq r_i$, -the repetition number of each $a_i$ is restricted by $r_i$, the item purchase restriction.

In these cases, the problem deals with the recognition of monotone functions, where:

(2) the zeros of the function appear in lower layers of the multivalued grid,

(3) the zeros of the function appear in some homogeneous bottom area of the grid.

In both cases it is more efficient to use the second algorithmic scheme of [6].

The idea is as follows. Let $F$ be a monotone function defined on $\Xi_{m+1}^n$. First we note that $N_F \cap \widehat{H}$ satisfies the monotonicity property, i.e., for arbitrary $a, b$ of $\widehat{H}$, if $a \geq b$ then $F(a) \geq F(b)$.

**Algorithm 2**

1. Firstly identify the part of the monotone function belonging to $\widehat{H}$ by one of the known resources of identification of monotone functions, and thus, reduce the size of the multi-valued grid. As a result we obtain $N_F \cap \widehat{H}$.
2. Apply the cube-splitting according to $N_F \cap \widehat{H}$, that considers the vertical equivalence classes $\mathcal{E}_1,\cdots,.\mathcal{E}_{|N_F \cap \widehat{H}|}$ only for the vertices of $N_F \cap \widehat{H}$.
3. Implement 2-4 Steps of Algorithm 1.

The algorithm can easily be adjusted for the identification of the complement of $F$ in $\Xi_{m+1}^n$.

## 4.2 Resources

To implement Step 1 of Algorithm 2 we have the following resources of identification of monotone functions in $\widehat{H}$.

a) The first resource is the known algorithm by V. Alexeyev [2]. We notice that applying the algorithm to identification of monotone functions on $\widehat{H}$ (instead of $\Xi_{m+1}^{n}$) is a significant reduction of the work. The complexity of the algorithm $U_0$ on $\widehat{H}$ will be:

$$T(U_0) \leq |M_1| + \left\lfloor log_2(\tfrac{m}{2}) \right\rfloor \cdot |N_1|,$$

where $M_1$ and $N_1$ are the middle layers of $\widehat{H}$, defined accordingly.

b) The second resource that we may use, is - applying the cube-splitting itself for identifying the function on $\widehat{H}$. We define the upper homogeneous area $\widehat{\widehat{H}}$ of $\widehat{H}$,: this is the set of all elements of $\Xi_{m+1}^{n}$ with all coordinate values $\geq 3m/4$; then

1. Apply the cube-splitting on $\widehat{H}$ and find $\mathcal{E}_1, \cdots, .\mathcal{E}_{\widehat{\widehat{H}}}$, equivalence classes according to the elements of $\widehat{\widehat{H}}$.
2. Compose the corresponding induced binary cube $E_i$ for every $\mathcal{E}_i$.
3. Apply the Hansel's algorithm to recognize the induced Boolean function $f_i$, defined on $E_i$ as follows: for every $\beta \in E_i$, $f_i(\beta) = 1$ if and only if $F(b) = 1$, where $b$ is the origin of $\beta$ in $\widehat{H}$ (for $i = 1, \cdots, \left|\widehat{\widehat{H}}\right|$).
4. Transfer the recognition results to $\widehat{H}$.

At this point we find $N_F \cap \widehat{H}$, and then continue with:

5. Apply the cube-splitting according to $N_F \cap \widehat{H}$, that is find the vertical equivalence classes $\mathcal{E}_1, \cdots, .\mathcal{E}_{|N_F \cap \widehat{H}|}$ only for the vertices of $N_F \cap \widehat{H}$.
6. Implement 2-4 Steps of Algorithm 1.

c) The cube-splitting can be applied recursively.

The following resource is also worth mentioning that can be used in all cases/algorithms. This is the growing technique [8] in monotone Boolean function recognition and chain computation algebra [9].

## 5. Conclusion

The problem of query based algorithmic recognition of monotone binary functions defined in multi-valued grids is known as a hard problem. It was investigated by V. Korobkov, V. Alekseev, A.Serjantov, and others. It is known a chain-split type algorithm developed by V. Alekseev, where the complexity estimates were given in terms of sizes of the middle layers of the grid. A novel method of identification of monotone binary functions based on the partitioning of the grid into discrete structures isomorphic to binary cubes was proposed in our recent work, where a theoretical level description of two algorithmic schemes was introduced. This paper provides the implementation details of the algorithms, as well as focuses on the recognition of monotone binary functions with a small number of units /or zeros/ distributed in homogeneous areas of the grid.

## Acknowledgement

This work is partially supported by the grant № 18T-1B407 of the Science Committee of the Ministry of Education and Science of Armenia.

# Մոնոտոն ֆունկցիայի ճանաչում բազմարժեք ցանցում

Լևոն Հ. Ասլանյան և Հասմիկ Ա. Սահակյան

ՀՀ ԳԱԱ Ինֆորմատիկայի և ավտոմատացման պրոբլեմների ինստիտուտ
e-mail: lasl@sci.am, hsahakyan@sci.am

## Ամփոփում

Այս շարքի նախորդ աշխատանքներում առաջարկվել է մոնոտոն բինար ֆունկցիայի վերծանման նոր մոտեցում՝ հիմնված բազմարժեք բազմաչափ ցանցի խորանարդատիպ տրոհման մեթոդի վրա, որտեղ տեսական մակարդակում առաջարկվել է խնդրի լուծման երկու ալգորիթմ /ալգորիթմական սխեմա/: Ներկա աշխատանքում տրվում են այդ ալգորիթմների իրականացման մանրամասները, ինչպես նաև դիտարկվում է այն դեպքը, երբ մոնոտոն ֆունկցիան ունի փոքր թվով մեկ արժեքի գագաթներ:

# Заметки о монотонном распознавании в многозначных решетках

Левон А. Асланян и Асмик А. Саакян

Институт проблем информатики и автоматизации НАН РА
e-mail: lasl@sci.am, hsahakyan@sci.am

## Аннотация

Новый подход монотонного распознавания на основе разбиения многозначной решетки на дискретные структуры, изоморфные бинарным кубам (метод “кубического разбиения”) предложен в серии последних работ, где на теоретическом уровне дано описание двух алгоритмов /алгоритмических схем/ решения задачи. В данной статье приводится подробное описание деталей реализации этих алгоритмов, а также рассматривается случай распознавания монотонной функции с небольшим числом единиц, что связано с рядом практических приложений.